\documentclass[runningheads]{svmult}

\usepackage{makeidx}   % allows index generation
\usepackage{graphicx}  % standard LaTeX graphics tool
\usepackage{subeqnar}  % subnumbers individual equations
\usepackage{multicol}  % used for the two-column index
\usepackage{physprbb}  % modified textarea for proceedings,
\makeindex             % used for the subject index
\catcode`\@=11
\def\gsim{\ifmmode{\mathrel{\mathpalette\@versim>}}
    \else{$\mathrel{\mathpalette\@versim>}$}\fi}
\def\lsim{\ifmmode{\mathrel{\mathpalette\@versim<}}
    \else{$\mathrel{\mathpalette\@versim<}$}\fi}
\def\@versim#1#2{\lower 2.9truept \vbox{\baselineskip 0pt \lineskip 
    0.5truept \ialign{$\m@th#1\hfil##\hfil$\crcr#2\crcr\sim\crcr}}}
\catcode`\@=12

\newcommand\msun{\rm M_\odot}
\newcommand{\Htwo}{{\rm H_2}}

\begin{document}
\title*{The Assembly of the First Galaxies}
\titlerunning{The Assembly of the First Galaxies}
\author{Zolt\'an Haiman\inst{1,2}}
\authorrunning{Z. Haiman}
% if there are more than two authors,
% please abbreviate author list for running head
%
\institute{Princeton University Observatory, Peyton Hall, 08544 NJ, USA
\and Hubble Fellow}
\maketitle              % typesets the title of the contribution

\begin{abstract}
The first galaxies formed at high redshifts, and were likely
substantially less massive than typical galaxies in the local
universe.  We argue that (1) the reionization of a clumpy
intergalactic medium (IGM) by redshift $z\approx 6$, (2) its
enrichment by metals by $z\approx 3$ without disturbing the Ly$\alpha$
forest, and (3) the presence of supermassive black holes powering the
recently discovered bright quasars at $z\sim 6$, strongly suggest that
a population of low--mass galaxies exists beyond redshifts $z\gsim 6$.
Although the first stars could have been born in dark matter halos
with virial temperatures as low as $T_{\rm vir}\approx 200$K,
collapsing as early as $z\sim 25$, the first galaxies likely appeared
in significant numbers only in halos with $T_{\rm vir}>10^4$K that
collapsed later ($z\sim 15$).  The gas in these more massive halos
initially contracts isothermally to high densities by atomic
Ly$\alpha$ cooling. ${\rm H_2}$ molecules can then form efficiently
via non--equilibrium gas--phase chemistry, allowing the gas to cool
further to $T\sim 100$K, and fragment on stellar mass scales.  These
halos can harbor the first generation of ``mini-galaxies'' that
reionized the universe.  The continuum and line emission from these
sources, as well as their Ly$\alpha$ cooling radiation, can be
detected in the future by {\it NGST} and other instruments.

\end{abstract}

\section{Introduction}
Recent measurements of the cosmic microwave background (CMB)
temperature anisotropies, determinations of the luminosity distance to
distant type Ia Supernovae (SNe), and other observations have led to
the emergence of a robust ``best--fit'' cosmological model with energy
densities in cold dark matter (CDM) and ``dark energy'' of
$(\Omega_{\rm M},\Omega_{\rm \Lambda})\approx (0.3,0.7)$.  The growth
of density fluctuations, and their evolution into non--linear dark
matter structures can be followed in detail from first principles by
semi--analytic methods~\cite{13,14} and N--body simulations~\cite{15}.
Structure formation is ``bottom--up'', with low--mass halos condensing
first. Halos with the masses of globular clusters, $10^{5-6}\msun$,
are predicted to have condensed as early as $\sim$1\% of the current
age of the universe, or redshift $z\sim 25$.  It is natural to
identify these condensations as the sites where the first
astrophysical objects, such as stars, or quasars, were born.

\section{Current Evidence for High Redshift Galaxies}

Current observations directly probe the universe out to redshift
$z\sim6$, with the record--holder quasar at $z=6.28$~\cite{16}, and
the history of star--formation and of quasar activity mapped out to
$z\sim 5$~\cite{17}.  However, there is convincing observational
evidence that an additional, yet undiscovered population of low--mass
galaxies exists at these redshifts and beyond.

%%%%%% Figure 1 %%%%%%
\begin{figure}[t]
\begin{center}
\includegraphics[width=0.8\textwidth]{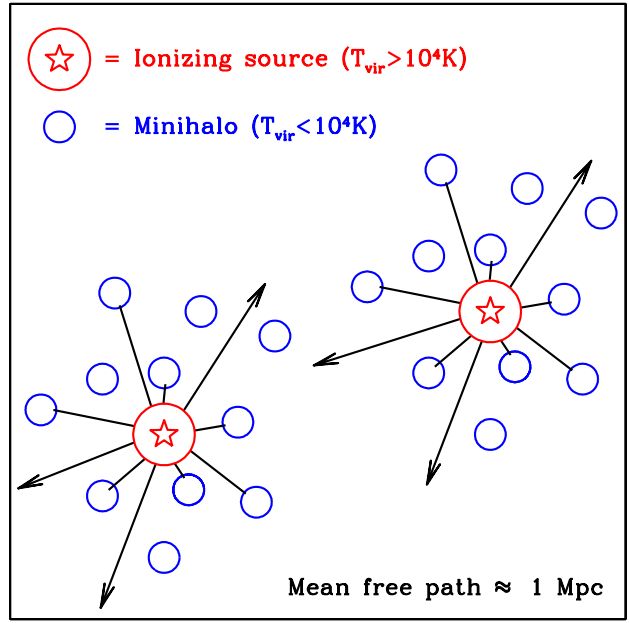}
\end{center}
\caption[]{Dense minihalos with virial temperatures 
$10^3$K$\lsim T_{\rm vir}\lsim 10^4$K are an important sink of
ionizing radiation as they surround UV sources at high redshifts
($z\gsim 6$).  As a population, such minihalos raise by about an order
of magnitude the budget of ionizing photons necessary to fully
reionize the universe by redshift $z\approx 6$~\cite{1}.}
\label{fig:sinks}
\end{figure}
%%%%%%%%%%%%%%%%%%%%%%%%

\subsection{Reionization of a Clumpy Universe}

The lack of a Gunn--Peterson trough in the spectra of all quasars to
date, except for the record holder quasar at $z=6.28$, imply that the
universe is highly ionized prior to redshift $z\sim 6$ (perhaps near
$z\sim 6.3$).  An early population of stars or quasars could have
photoionized the IGM, if they produced at least one ionizing photon
per hydrogen atom in the universe. A naive extrapolation of the
luminosity density of bright quasars towards $z=6$ reveals that these
sources fall short of this requirement.  Extrapolating the known
population of Lyman Break Galaxies (LBGs) towards $z=6$ comes closer:
assuming that 15$\%$ of the ionizing radiation from LBGs escapes into
the IGM (on average, relative to the escape fraction at 1500\AA), a
naive extrapolation shows that LBGs emitted $\sim$ one ionizing photon
per hydrogen atom prior to $z=6$~\cite{1}.  Although this would be
sufficient to ionize every H atom once, the required photon budget
exceeds this value.  The earliest ionizing sources are likely
surrounded by numerous ``minihalos'' that had collapsed earlier, but
had failed to cool and form any stars or quasars\footnote{If minihalos
are themselves the sources of ionizing radiation, and their gas is
photo--evaporated ``inside--out'', then the required photon budget is
likely to be even higher.}.  This is illustrated in
Figure~\ref{fig:sinks}.  The minihalos have typical masses below
$10^7\msun$, and represent a population of dense clumps that is
currently unresolved in large three--dimensional cosmological
simulations.  The UV radiation incident on the minihalos heats their
gas to $T\approx10^4$K, causing it to photo--evaporate~\cite{19,12}.
The mean free path of ionizing photons, before they are absorbed by an
evaporating minihalo, is about $\sim 1$ (comoving) Mpc. As a result,
the typical fate of an ionizing photon, emitted at $z\gsim 6$, is to
be absorbed by a minihalo within a small fraction of the Hubble
distance -- before it could contribute to the reionization of the bulk
of the IGM.  A simple model of the photoevaporation process~\cite{1},
summed over the expected population of minihalos, reveals that on
average, an H atom in the universe recombines $\gsim 10$ times before
redshift $z=6$.  The implication is that the ionizing emissivity at
$z>6$ was $\sim10$ times higher than provided by a straightforward
extrapolation back in time of known quasar and galaxy populations.

\subsection{Metal Enrichment of the Intergalactic Medium}

\begin{figure}[t]
\begin{center}
\includegraphics[width=0.8\textwidth]{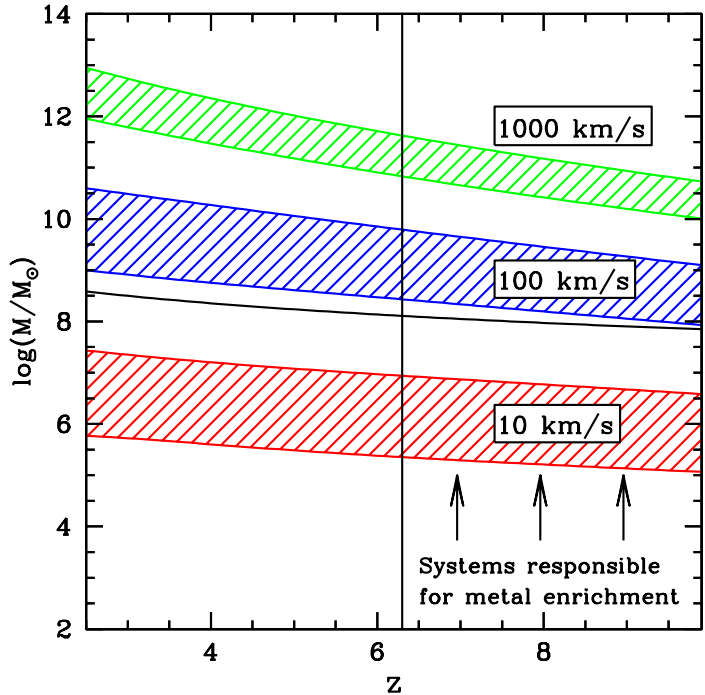}
\end{center}
\caption[]{Typical masses of halos responsible for dispersing metals into 
the IGM if metals travel at mean speeds of 10, 100, or 1000 km/s.  The
upper (lower) envelope for each shaded curve corresponds to the
requirement that 3 (100)\% of the volume is enriched within the age of
the universe at redshift $z$.  The solid horizontal curve shows the
halo that has a velocity dispersion of $\sigma=10$km/s; the vertical
line shows the minimum reionization redshift at $z=6.3$.  Widely
dispersing the metals at low ($\sim 10$km/s) speeds requires
enrichment by low--mass systems, $M\approx 10^{6}\msun$, which produce
metals only prior to reionization.}
\label{fig:metals}
\end{figure}

Recent detections of CIV and SIV absorption associated with low column
density Ly$\alpha$ absorption lines in the spectra of distant quasars
imply that the universe was enriched by metals to a mean level of
$\sim 10^{-3} Z_\odot$ prior to redshift $z\sim 3$.  These absorbers,
whose HI column densities are as low as $10^{14.2}~{\rm
cm^{-2}}$~\cite{20}, are identified in cosmological hydrodynamical
simulations as regions with typical densities only a few times above
that of the mean IGM.  This low density rules out ``in--situ'' metal
enrichment, and raises the question: where did these metals come from?
LBGs discovered at redshifts $z=3-4$ would be natural candidates for
producing and dispersing heavy elements in galactic winds~\cite{7}.
In order for a significant fraction of the volume to be enriched
within a Hubble time, the winds need to move heavy elements at mean
speeds close to $1000$ km/s (the typical separation of LBGs is a few
Mpc).  While outflows at such high speeds are known to occur, it is
unclear whether this scenario can be reconciled with the observed
line--widths of the absorbers, which are as narrow as $\sim 20$ km/s,
when the cooling time in the low--density IGM is exceedingly long at
$z\sim 3$.
An alternative possibility is early enrichment by much more numerous
low--mass systems, with typical separations of $<0.1$Mpc, which can
drive outflows at their escape velocities of a few$\times 10$ km/s
(see also~\cite{2}).  In Figure~\ref{fig:metals}, we show the maximum
masses of halos (using the halo mass function from~\cite{15}) whose
mean separation is small enough so that their distance can be crossed
in the age of the universe (at each redshift) at speeds of 10, 100, or
1000 km/s.  In all three cases, the shaded regions correspond to
requiring that the dispersed metals fill 3-100\% of the volume.  The
figure reveals that at $z\approx 3-4$, systems with $M\approx
10^{12}\msun$ are sparse, and require speeds of $\sim 1000$km/s.  Only
systems with masses as low as $M\approx 10^{6-7}\msun$ (lowest shaded
curve) are sufficiently abundant to widely disperse metals with
outflow speeds of a few $\times 10$ km/s.  However, such low--mass
systems have velocity dispersions below $\sigma=10$km/s, and are
unlikely to have formed metals in the presence of the
post--reionization UV background (the horizontal solid curve shows the
halo mass corresponding to $\sigma=10$km/s).  The natural conclusion
is that the metals had to be produced and dispersed by low--mass
systems, $M\approx 10^{6}\msun$, prior to reionization at $z\gsim
6.3$.

%%%%%% Figure 2 %%%%%%
%%%%%%%%%%%%%%%%%%%%%%%%

\subsection{The Growth of Supermassive Black Holes}

A third line of evidence for high--redshift activity comes from the
sheer size of supermassive black holes (BHs) required to power the
recently discovered bright quasars near $z\approx 6$.  Assuming that
these quasars are shining at their Eddington limit, and are not beamed
or lensed\footnote{Strong lensing or beaming would contradict the
large proximity effect around these quasars~\cite{21}.}, their BH
masses are inferred to be $M_{\rm bh}\sim 4\times10^9\msun$.  The
Eddington--limited growth of these supermassive BHs by gas accretion
onto stellar--mass seed holes, with a radiative efficiency of
$\epsilon\equiv L/\dot mc^2\approx 10\%$, requires $\sim 20$
e--foldings on a timescale of $t_{\rm E}\sim
4\times10^7(\epsilon/0.1)$ years.  While the age of the universe
leaves just enough time ($\lsim 10^9$ years) to accomplish this growth
by redshift $z=6$, it does mean that accretion has to start early, and
the seeds for the accretion have to be present at ultra--high
redshifts: $z\gsim 15 (20)$ for an initial seed mass of
$100(10)\msun$. Furthermore, the radiative efficiency cannot be much
higher than $\epsilon\approx 10\%$~\cite{22,23}.  Since an individual
quasar BH could have accreted exceptionally fast (exceeding the
Eddington limit), it will be important to apply this argument to a
larger sample of high--redshift quasars. Nevertheless, we note that a
comparison of the light output of quasars at the peak of their
activity ($z\sim 2.5$) and the total masses of their remnant BHs at
$z=0$ shows that during the growth of most of the BH mass the
radiative efficiency cannot be much smaller than $10\%$, and hence any
'super-Eddington' phase must be typically restricted to building only
a small fraction of the final BH mass.

\section{The Assembly of the First Galaxies}

While the formation of non--linear dark matter halos can be followed
from first--principles, the formation of stars or BHs in these halos
is much more difficult to model ab--initio. Nevertheless, we may
identify three important mass--scales, which collapse at successively
smaller redshifts,
\begin{enumerate}
\item Gas can only condense in dark halos above the cosmological Jeans 
mass, $M_{\rm J}\approx 10^4\msun [(1+z)/11]^{3/2}$, so that the gravity of dark
matter can overwhelm gas pressure in the IGM.
\item Gas that condensed into Jeans--unstable halos can cool and contract
further in halos with masses above $M_{\rm H2} \gsim 10^5\msun[(1+z)/11]^{-3/2}$
(virial temperatures of $T_{\rm vir}\gsim 10^2$K), provided there is
a sufficient abundance of ${\rm H_2}$ molecules (at a level of at least
$n_{\rm H2}/n_{\rm H}\sim 10^{-3}$).
\item In halos with masses above $M_{\rm H} \gsim 10^8\msun[(1+z)/11]^{-3/2}$
(virial temperatures of $T_{\rm vir}\gsim 10^4$K), gas can cool and
contract via excitation of atomic Ly$\alpha$, even in the absence of
any ${\rm H_2}$.
\end{enumerate}

The first stars or BHs likely formed in gas that cooled and condensed
via excitations of roto-vibrational levels of ${\rm H_2}$ molecules in
dark matter condensations with $M\sim10^5\msun$ at redshift $z\sim 25$
(corresponding to $\sim$3$\sigma$ peaks in the primordial density
field)~\cite{5,6}.  Unless the first rare objects were significant
sources of X--ray photons with energies $E\gsim 1$keV, their soft UV
radiation, permeating the distant universe at photon energies
$E<13.6$eV, caused a negative feedback, strongly suppressing ${\rm
H_2}$ cooling and star--formation in clumps that collapsed
subsequently~\cite{3,8}. Efficient and widespread star (and/or BH)
formation, capable of reionizing the universe, had to then await the
collapse of halos with $T_{\rm vir} > 10^4$K, or $M_{\rm halo} > 10^8
[(1+z)/11]^{-3/2} \, {\rm M_\odot}$.

%%%%%% Figure 3 %%%%%%
\begin{figure}[t]
\begin{center}
\includegraphics[width=0.8\textwidth]{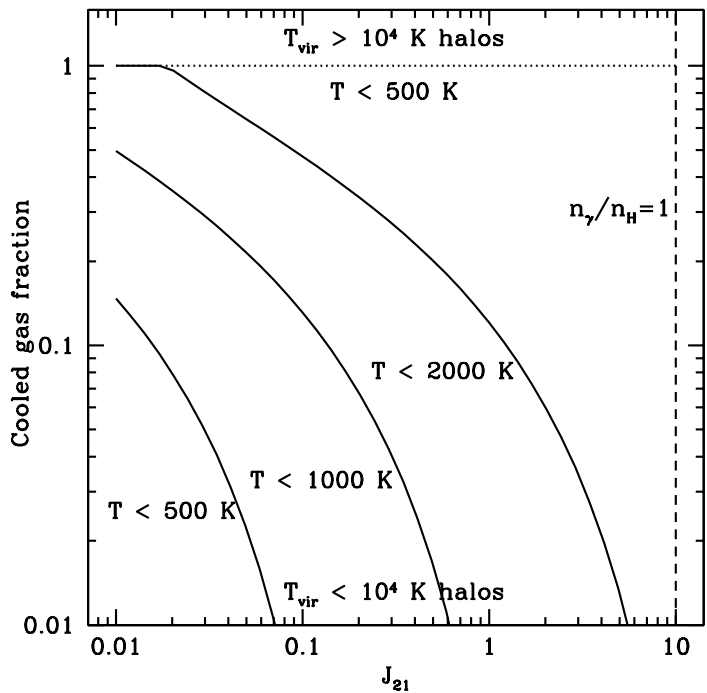}
\end{center}
\caption[]{The fraction of gas in a halo that can cool below a given 
temperature $T$, as a function of the external UV radiation field
$J_{21}$, for halos at z=15. In halos with $T_{\rm vir} < 10^{4}$K
(lower three solid curves), only a small fraction of the gas is at
sufficiently high density to be unaffected by external radiation
fields and cool. By contrast, in halos with $T_{\rm vir} > 10^{4}$K,
where atomic cooling operates, the gas contracts to high densities,
and virtually all of the disk gas can form ${\rm H_2}$ and cool to low
temperatures (dotted horizontal line). The UV intensity corresponding
to one ionizing photon per baryon in the universe is marked by the
dashed vertical line on the right. Adopted from~\cite{4}.}
\label{fig:fcool}
\end{figure}
%%%%%%%%%%%%%%%%%%%%%%%%

The evolution of halos with $T_{\rm vir} > 10^{4}$K differs from their
less massive counterparts~\cite{4}.  Efficient atomic line radiation
allows rapid cooling to $\sim 8000$ K; subsequently the gas can
contract to high densities nearly isothermally at this temperature.
In the absence of ${\rm H_2}$ molecules, the gas would likely settle
into a locally stable disk and only disks with unusually low spin
would be unstable.  However, the initial atomic line cooling leaves a
large, out--of--equilibrium residual free electron fraction. This
allows the molecular fraction to build up to a universal value of
$x_{\rm H_2}\approx10^{-3}$, almost independently of initial density
and temperature (this is a non--equilibrium freeze--out value that can
be understood in terms of timescale arguments~\cite{4}).

Unlike in less massive halos, $\Htwo$ formation and cooling is much
less susceptible to feedback from external UV fields. This is because
the high densities $n$ that can be reached via atomic cooling. The
${\rm H_2}$ abundance that can build up in the presence of a UV
radiation field $J_{21}$, and hence the temperature to which the gas
will cool, is controlled by the ratio $J_{21}/n$.  For example, in
order for a parcel of gas to cool down to a temperature of 500K, this
ratio has to be less than $\sim 10^{-3}$ (where $J_{21}$ has units of
$10^{-21} {\rm erg \, s^{-1} \, cm^{-2} Hz^{-1} \, sr^{-1}}$, and $n$
has units of ${\rm cm^{-3}}$). In Figure~\ref{fig:fcool}, we show, as
a function of the external UV radiation field $J_{21}$, the mass
fraction of gas which is able to cool to a temperature $T$ in $T_{\rm
vir} < 10^{4}$K halos (the gas is assumed to be in hydrostatic
equilibrium within an NFW halo~\cite{25}), and $T_{\rm vir} > 10^{4}$K
halos (the gas is assumed to have cooled via atomic Ly$\alpha$ to a
$10^4$K, rotationally supported disk~\cite{26}).

The figure reveals that flux levels well below that required to fully
reionize the universe strongly suppresses the cold gas fraction in
$T_{\rm vir} < 10^{4}$K halos.  By comparison, the UV flux has nearly
negligible impact on ${\rm H_2}$ formation and cooling in $T_{\rm vir}
> 10^{4}$K halos, where all of the gas is able to cool to $T=500$K.
Indeed, under realistic assumptions, the newly formed molecules in the
dense disk can cool the gas to $\sim 100$ K, and allow the gas to
fragment on scales of a few $\times 100$~M$_\odot$. Various feedback
effects, such as $\Htwo$ photodissociation from internal UV fields,
and radiation pressure due to Ly$\alpha$ photon trapping, are then
likely to regulate the eventual efficiency of star formation in these
systems.

\section{Future Observational Signatures}

Although the first generation of galaxies are distant, and
intrinsically faint objects, they should be within reach of {\it
NGST}.  If $\sim 10\%$ of the gas turns into stars in a $10^8\msun$
halo, it should be detectable in the $1-5\mu$m band with {\it NGST}
out to redshifts beyond $z=10$ in a $10^4$ second integration.  A
$\sim10^5\msun$ black hole, forming out of $\sim1\%$ of the gas in
such a halo, and shining at its Eddington limit, would be detectable
to a similar redshift.  Simple ``semi--analytical''
models~\cite{24,18} predict that {\it NGST} will either detect a
significant number of ultra--high redshift sources, or else it will
severely constraint any early model of early structure formation.
Similarly optimistic conclusions can be drawn about detecting
recombinant line emission from high--redshift sources: star formation
rates as small as $\sim 1\msun$/yr, or BHs as small as $\sim10^5\msun$
translate into detectable H$\alpha$ and He line fluxes beyond $z=10$,
as long as the escape fraction of H-- and He--ionizing radiation from
these sources is low~\cite{11}. Finally, as the baryons cool and
contract inside high--redshift halos with virial temperatures
$T\gsim10^4$K, they likely channel a significant fraction of their
gravitational binding energy into the Ly$\alpha$ line.  At the
limiting line flux $\approx 10^{-19}~{\rm
erg~s^{-1}~cm^{-2}~asec^{-2}}$ of the {\it NGST}, several sufficiently
massive halos, with velocity dispersions $\sigma\gsim 120~{\rm
km~s^{-1}}$, would be visible per $4^\prime\times4^\prime$ field.  The
halos would have characteristic angular sizes of $\sim 10$\H{}, and
could be detectable in a broad--band survey out to $z\approx 6-8$ (but
not beyond the reionization redshift, where Ly$\alpha$ photons are be
resonantly scattered).  Their detection would provide a novel and
direct probe of galaxies caught in the process of their
formation~\cite{9,10} -- possibly before the first stars or quasars
even lit up.

\vspace{\baselineskip}
I thank the organizers of this workshop for their kind invitation, and
Peng Oh for many recent fruitful discussions, and for permission to
draw on our joint work.  I acknowledge support from NASA through a
Hubble Fellowship.

\end{document}